\documentstyle[12pt,epsfig,cmb]{article}

\newcommand{\preprintno}[2]{
{\begin{flushright} {#1}\\{#2}\end{flushright}}}

\renewcommand{\heading}[1]{\vspace*{5mm}
{\Large\begin{center} {\bf{#1}} \end{center}}}
\renewcommand{\author}[3]{\vspace{5mm}  
\begin{center}
{\normalsize \rm #1}\\    
{\normalsize \it #2}\\    
{\normalsize \it #3}\\    
\vspace{0.65cm}\makebox[3.8truecm]
{
\epsfxsize=3.8truecm
\epsfbox{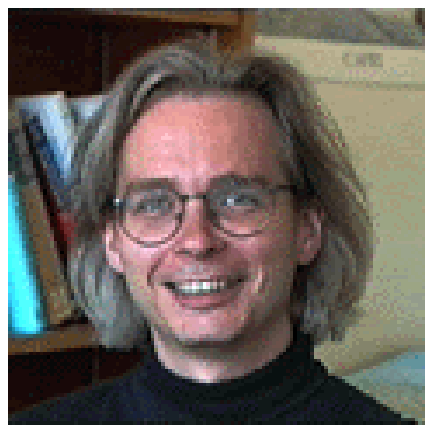}
}
\vspace{0.35cm}\end{center} }

\def\ga{\gamma}
\def\de{\delta}

\def\De{\Delta}

\def\Th{\Theta}

\newcommand{\dc}{\delta_c}
\newcommand{\dr}{\delta_\gamma}
\newcommand{\dzz}{\delta_{00}}
\newcommand{\h}{\frac{\dot a}{a}\,}
\newcommand{\bv}{\mbox{\boldmath $v$}}
\newcommand{\bea}{\begin{eqnarray}}
\newcommand{\eea}{\end{eqnarray}}
\newcommand{\ben}{\begin{equation}}
\newcommand{\een}{\end{equation}}
\def\TTeng{\langle {\Theta_{00}(k,\eta)\Theta_{00}^*(k,\eta')\rangle}}
\def\TTmom{\langle {U(k,\eta)U^*(k,\eta')\rangle}}
\def\TTcross{\langle {U(k,\eta)\Theta_{00}^*(k,\eta')\rangle}}

\begin{document}

\preprintno{SUSX-TH-96-010}{{\tt astro-ph/9607115}}
\heading{COSMIC STRINGS AND COHERENCE}

\author{M.Hindmarsh $^{1}$, M.Sakellariadou $^{2}$, G.Vincent $^{1}$} 
{$^{1}$ School of Mathematical and Physical Sciences, University of Sussex, 
Brighton BN1 9QH, U.K.}  {$^{2}$ D\'epartement de Physique Th\'eorique,
Universit\'e de Gen\`eve,
Quai Ernest-Ansermet 24,
CH-1211 Gen\`eve 4,
Switzerland}

\begin{abstract}{\baselineskip 0.4cm 
Cosmic strings provide a radically different paradigm for the 
formation of structure to the prevailing inflationary one.  They 
afford some extra technical complications: 
for example, the calculation of the power spectrum of matter 
and radiation perturbations requires the knowledge of the history of 
the evolution of the defects in the form of two-time correlation 
functions. We describe some numerical simulations 
of string networks, designed to measure the two-time 
correlations during their evolution.  
}
\end{abstract}

\section{Introduction}

Calculations of the Cosmic Microwave Background (CMB) 
fluctuations from cosmic strings
\cite{HinKib95} have proved 
much harder than their inflationary equivalents.  One of the reasons is 
that the energy-momentum of the defects cannot be considered to be 
a perfect fluid like most of the rest of the contents of the Universe: 
instead we have to use numerical simulations to find how important 
quantities such as the energy density and velocity evolve.  The defect 
energy-momentum is then a
source for perturbations in the gravitational field, which must be 
calculated either using Greens functions \cite{VeeSte90} or 
by direct numerical integration \cite{PenSpeTur94}.

A major problem with using numerical simulations is the lack of dynamic 
range, even with supercomputers.  The largest simulations can only 
span a ratio of conformal times of about 10, which is not totally satisfactory. 
For calculating simple quantities such as power spectra  an 
alternative approach is available:  to use numerical simulations to 
construct an accurate model of the appropriate source functions, 
which are two-time correlation functions of various components of the 
energy-momentum tensor.  Relatively simple Greens functions can then 
be used to compute the power spectra of perturbations in perfect fluids,
including CDM and tightly coupled photons and baryons.

This talk summarises some work in progress \cite{VinHinSak96} 
on the properties of 
cosmic strings as sources of the gravitational 
perturbations.  An issue of interest which we address is that 
of {\em coherence}, recently raised in \cite{Alb+96}, and explored 
in more detail by the same authors in \cite{Mag+96b}. 

\section{Perturbations from defects}
A simple model serves to illustrate the issues involved.  We suppose 
that the Universe is spatially flat with zero cosmological constant, and 
consists of Cold Dark Matter (CDM), photons, baryons, and defects, with 
average densities $\rho_c$, $\rho_\ga$, $\rho_B$ and $\rho_s$ 
respectively. 
  The fluid density perturbations are written 
$\de_c$, $\de_\gamma$, $\de_B$ and the velocity perturbations are 
$\bv_c$, $\bv_\gamma$, $\bv_B$. 
The defect energy-momentum tensor is $\Th_{\mu\nu}$, which is also a 
small perturbation to the background, and to first order can 
be considered separately conserved (or {\em stiff} \cite{VeeSte90}). 
We take the synchronous gauge. 
Before decoupling at a redshift of $z_{\rm rec} \simeq 1100$ 
(assuming the standard ionization history), the photons and baryons 
were tightly coupled by Thompson scattering, which forces 
$\de_B = \frac{3}{4}\de_\gamma$ and $\bv_B = \bv_\gamma$. 

It is very convenient to use an entropy 
$s = (\frac{3}{4} \dr - \dc)$, and its time derivative $\dot s$, which is 
equal to the divergence of the radiation peculiar velocity.  
It is also convenient 
to introduce the pseudoenergy $\tau_{00}$ \cite{VeeSte90}, which is 
part of an ordinarily conserved energy-momentum pseudotensor:
\ben
\tau_{00} = \Theta_{00} + \frac{3}{8\pi G} \left(\h\right)^2 
(\Omega_c \dc + \Omega_\gamma(1+R) \dr)  + 
\frac{1}{4\pi G} \left(\h\right) \dot \dc,
\een
where $R=3\rho_B/4\rho_\gamma$.
In the sychronous gauge the pseudo-energy is in fact proportional 
to the scalar curvature of the 
constant conformal time slices.

Let us now define the pseudo-energy perturbation 
\ben
\dzz = \frac{8\pi G}{3} \left(\frac{a\,}{\dot a}\right)^2\tau_{00}.
\een
We may use the conservation equations for the defects to write 
the equations of motion of the fluids as four 
first-order equations \cite{PenSpeTur94}
\bea
\dot \dc & = &  \frac{3}{2} \left(\h\right) 
[\dzz - (\Omega_c + 2 \Omega_\gamma(1+R))\dc - 2\Omega_\gamma(1+R) s] - 
\mbox{\boldmath${\displaystyle 
\frac{3}{2} \left(\h\right)\Omega_s \de_s}$},\\
\dot s & = & y,\\
\dot \dzz & = & \left( \h \right) (1+3w)\dzz + \frac{4}{3} \Omega_\gamma 
(1+R) y + \mbox{\boldmath${\displaystyle\Omega_s kv_s}$},\\
\dot y & = & - k^2 c_s^2 (s + \dc) - \left(\h\right) 
(1 - 3c_s^2) y,
\eea
where $w = p/\rho$, the ratio of the total pressure to the total 
energy density, and $c_s^2 = 1/3(1+R)$.

These equations have the inhomogeneous form $\dot \Delta_i = M_i^j
\Delta_j + \Sigma_i$, 
where $\Delta = (\dc,s,\dzz,y)^T$, and the source vector, picked 
out in bold above, is 
$\Sigma = (-\frac{3}{2} (\dot a/a)\Omega_s \de_s,\; 0,\; 
\Omega_s kv_s,\; 0)^T$.
The solution to these equations with 
initial condition $\Delta(\eta_i)$ is then
\ben
\Delta(\eta) = G(\eta,\eta_i) \Delta(\eta_i) + \int_{\eta_i}^\eta 
\,d\eta' G(\eta,\eta') \Sigma(\eta'),
\een
where $G(\eta,\eta')$ is the Green's function.
The initial perturbation is $\Delta(\eta_i) = (\dc(\eta_i),0,0,0)^T$, 
corresponding to perturbations which are both adiabatic, in the sense that 
$s=0$, and isocurvature, in the sense that $\tau_{00} = 0$. In the presence 
of defects, the isocurvature condition forces the initial perturbations in 
the matter and radiation to {\em compensate} those of the defects.   
However, the initial 
compensation can be ignored, as the initial condition 
consists only of decaying modes, which
is one of the advantages of using this basis for the perturbations 
\cite{PenSpeTur94}.
An initial condition consisting purely 
of adiabatic growing modes  would be 
$\Delta(\eta_i) = (\dc(\eta_i),0,2\dc(\eta_i),0)^T$.

At time $\eta$, the power spectra of the various 
perturbations are therefore 
\ben
\langle|\Delta_i(\eta)|^2\rangle = 
\int_{\eta_i}^\eta \,d\eta_1 \int_{\eta_i}^\eta\,d\eta_2
G_i^j(\eta,\eta_1) G_i^k(\eta,\eta_2) \langle \Sigma_j(\eta_1) 
\Sigma_k^*(\eta_2) \rangle,
\label{e:PS}
\een
(no sum on $i$).  It is clear from this equation that the calculation 
of the power spectra requires the knowledge of the two-time 
correlation functions
\ben
C^{\rho\rho}(\eta_1,\eta_2) = \langle \delta_s(\eta_1) \de_s^*(\eta_2)\rangle, 
\quad
C^{UU}(\eta_1,\eta_2) = \langle v_s(\eta_1) v_s^*(\eta_2)\rangle, 
\quad
X^{\rho U}(\eta_1,\eta_2) = \langle \delta_s(\eta_1) v_s^*(\eta_2)\rangle,
\label{e:corrs}
\een
where the angle brackets are to be understood as averages over ensembles 
of topological defects.

\section{Flat space cosmic string simulations}

In order to compute the averages in (\ref{e:corrs}) 
we performed many simulations of networks of string in 
Minkowski space \cite{SmiVil87,SakVil88}.  
It is not at first sight such a good idea to abandon the expanding Universe, 
but there are arguments which support the procedure. Firstly, a spatially 
flat FRW cosmology is conformal to Minkowksi space, so one can identify  
Minkowksi time with conformal time, and Minkowksi space coordinates 
with comoving space coordinates. Secondly, the string network scale length
is significantly smaller than the horizon length, which means that 
the effects of the expanding background on the dynamics of the 
string should be small.  Thirdly, a string network can be well described
by a couple of parameters:  the string density and the correlation 
length of the network \cite{CopKibAus92}, which means we can hope to 
translate Minkowksi space results into FRW results by appropriate 
scalings.  Lastly, the savings in computational resources are 
immense. The strings can be evolved on a cubic spatial lattice, which means
that only integer arithmetic need be used, and that self-intersections 
are extremely easy to check for.

A vital property possessed by string networks is known as {\em scaling}. 
In its simplest form, scaling means that all quantities with 
dimensions of length are proportional to a fundamental network scale 
$\xi$, which is in turn a function of length scales in the 
network dynamics.  In a cosmological setting, the only scale in the 
equations of motion is the horizon or the conformal time $\eta$. 
Thus $\xi \propto \eta$, where $\xi$ is understood to be a comoving scale. 
The invariant length density $\L$ (which is proportional to the energy density 
through the string mass per unit length $\mu$) is therefore, on 
naive dimensional grounds, proportional to $\xi^{-2}$. In fact, it is 
convenient to {\em define} $\xi$ by $\xi = 1/\sqrt{\L}$.
The simulations conserve energy and thus invariant length. 
However, real strings decay into particles and gravitational radiation, 
so we model this by excluding small loops below a threshold $l_c$ from 
being counted as string.  This threshold may be chosen in many ways: 
it can be a constant, or it may be a fixed fraction of $\xi$.  However, 
we find that the remainder (``long'' or ``infinite'' string) always scales 
towards the end of the simulations, that is, $\xi=x\eta$, 
although the constant $x$ depends on how the cut-off is chosen.

Scaling has important implications for the correlation functions.  
For example, the correct scaling behaviour for the power spectrum of 
the string density is
\ben
C^{\rho\rho}(\eta,\eta) = \frac{1}{\xi} P^{\rho}(k\xi).
\een
We can also express the velocity power spectrum in terms of 
a similar scaling function $P^{U}(k\eta)$.
These forms ensure that the mean square fluctuation in the string density 
at horizon crossing is constant, when $\xi/\eta$ is constant. 
Hence the compensating 
fluctuations in the matter and radiation are also constant at horizon 
crossing, which gives a Harrison-Zel'dovich spectrum.
\begin{figure}
\centerline{\epsfig{file=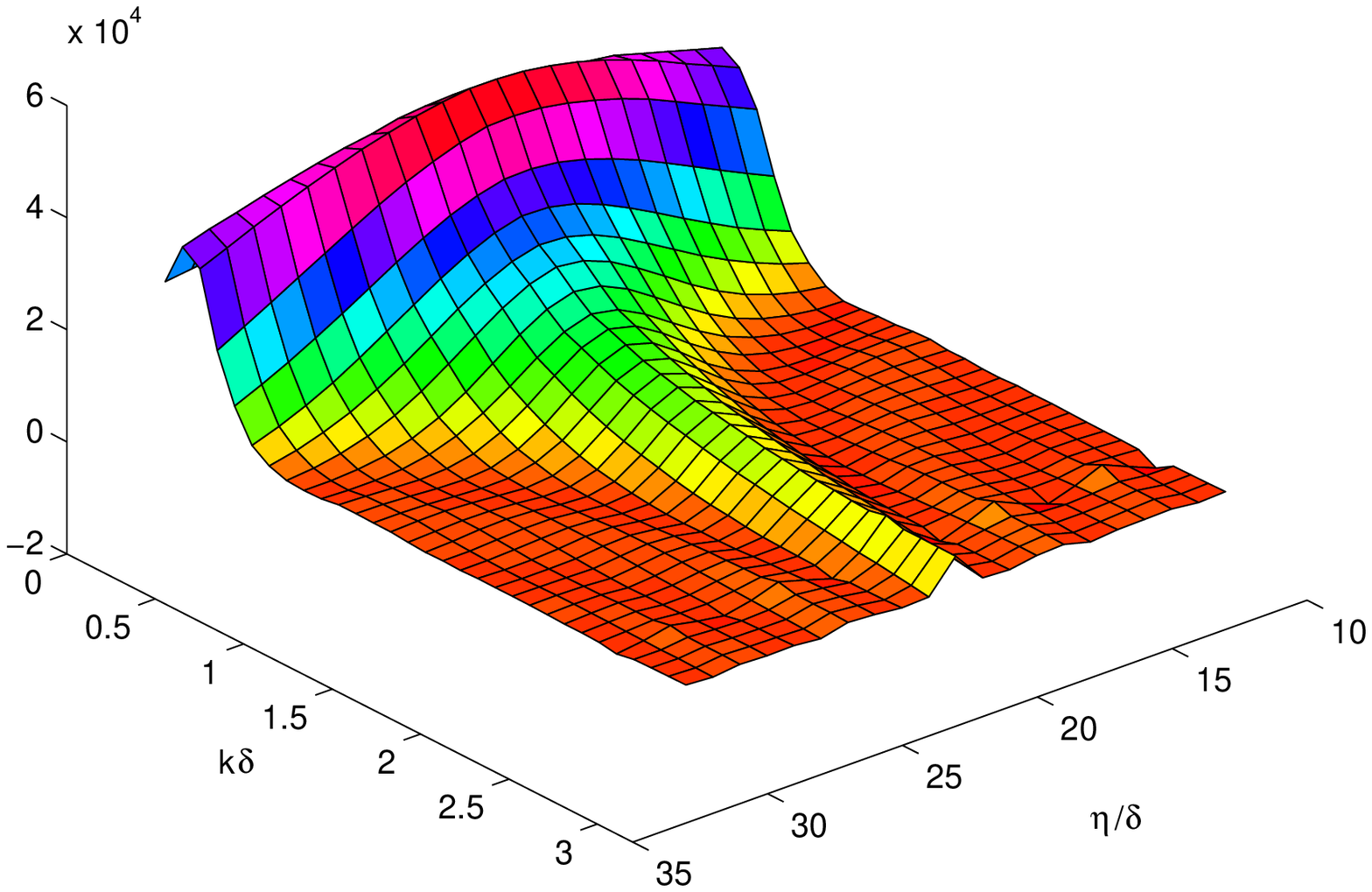,width=7.5cm,angle=0}}
\caption{Two-time correlation function $\TTeng$ for 
$\eta'=22\delta$}
\label{fig:ttcf00}
\end{figure}

In Figures 1, 2, and 3 the correlation functions (\ref{e:corrs}) measured 
for strings whose length is above the threshold $l_c$ 
are displayed.  
They result from many simulations on $(64\delta)^3$ or 
$(128\delta)^3$ lattices, where $\de$ is the lattice spacing, with
approximately $7500$ or $60000$ string points. 
For calculating the two-time correlation functions we typically
averaged over 50 simulations.\footnote{For more details of the simulations, 
the reader is referred to \cite{VinHinSak96}.}  We find that good fits in the 
important regions where the correlators are large are obtained from 
the following functions
\bea
 C^{\rho\rho}&=&{1 \over \sqrt {\xi\xi'}} 
\sqrt{P^{\rho}(k\xi)P^{\rho}(k\xi')}
  {\it e}^{-{\frac{1}{2}} \Upsilon^2 k^2 (1-(k\Delta)) (\eta-\eta')^2}
 \label{2t00approx}
\\ 
C^{\rho U}&=&{1 \over \sqrt {\xi\xi'}} 
\sqrt{P^{\rho}(k\xi)P^{\rho}(k\xi')}
 {\it e}^{-{\frac{1}{2}} \Upsilon'^2 k^2 (\eta-\eta')^2} ({\alpha \over 
k\sqrt{\xi\xi'}}-\Upsilon'^2 k (\eta-\eta'))
 \label{2tcrossapprox}
\\
 C^{UU}&=&{1 \over \sqrt {\xi\xi'}} \sqrt{P^{U}(k\xi)P^{U}(k\xi')}
 {\it e}^{-{\frac{1}{2}} \Upsilon''^2 k^2 (\eta-\eta')^2} (1-\Upsilon''^2 
k^2 
(\eta-\eta')^2)
 \label{2t0iapprox}
 \eea
The values for $\Upsilon$, $\Upsilon'$ and $\Upsilon''$  are 
given in 
Table \ref{tab1}.
 $\Delta$ is approximately $3\delta$, which indicates that it is 
 probably a lattice artifact.
\begin{center}
{{\bf Table 1.} The values of the parameters for the models in equations 
(\ref{2t00approx}),
(\ref{2tcrossapprox})  and (\ref{2t0iapprox}), obtained by 
minimising the $\chi^2$.} 
\end{center}
\begin {center}
\begin {tabular} {|c|c|c|c|c|}
\hline
$l_c$ & $\Upsilon$ & $\alpha$ & $\Upsilon'$& $\Upsilon''$\\
\hline
$2\delta$ & $0.21 \pm 0.05$ & $0.19 \pm 0.05$ & $0.42 \pm 0.05 $ & 
$0.36 \pm 0.07$\\
$4\delta$ & $0.18 \pm  0.05$  & $0.16 \pm 0.05$ & $0.42 \pm 0.05 $ 
& 
$0.42 \pm 0.06$\\
\hline
\end {tabular}

\label{tab1}
\end{center}
\begin{figure}
\centerline{\epsfig{file=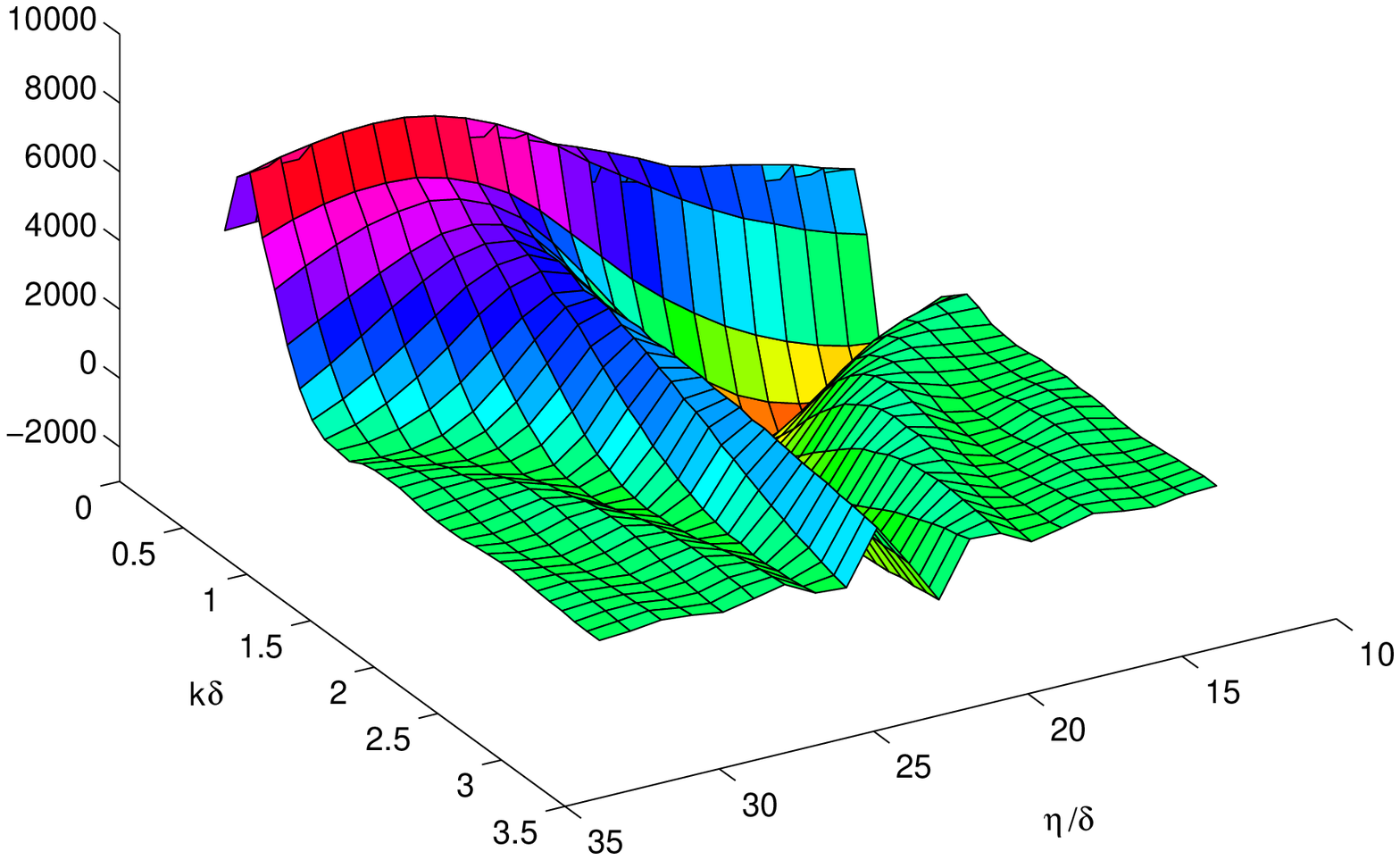,width=7.5cm,angle=0}}
\caption{Two-time correlation function $\TTcross$ for 
$\eta'=22\delta$}
\label{fig:ttcfcross}
\end{figure} 

\section{Implications for strings in FRW backgrounds}

The effect of the expansion of the Universe in a Friedmann model is 
to reduce the velocity string segments by Hubble damping, just as for 
particles.  However, as the string moves relativistically, the 
equation of motion becomes non-linear.  The 
conservation equations for $\Theta_{00}$ and $U = i\hat k_i\Th_{0i}$ 
become 
\bea
\dot \Theta_{00} + \h (\Th_{00}+\Th)+kU &=& 0,\label{e:EMden}\\
\dot U + 2\h U +\frac{1}{3} k (\Th + 2 \Th^{\rm s}) & = & 0, 
\eea
where $\Th \equiv \Th_{ii}$ and $\Th^{\rm s} \equiv (\hat{k}_i\hat{k}_j 
-\frac{1}{3}\de_{ij})\Th_{ij}$.
Both $\Th$ and $\Th^{\rm s}$ are unconstrained by energy-momentum
conservation, and so one could imagine that fluctuations in the pressure 
could drive extra fluctuations in the energy density 
through the second term in (\ref{e:EMden}).  As the scale of this term 
is $\eta^{-1}$ rather than $k$, this could change the 
conclusions of the previous section, that the coherence time for a 
mode of wavenumber $k$ was proportional to $k^{-1}$.  Instead we could 
have the situation assumed in \cite{Alb+96}, with coherence 
time proportional to $\eta$.  Outside the horizon, this makes a significant 
difference.
\begin{figure}
\centerline{\epsfig{file=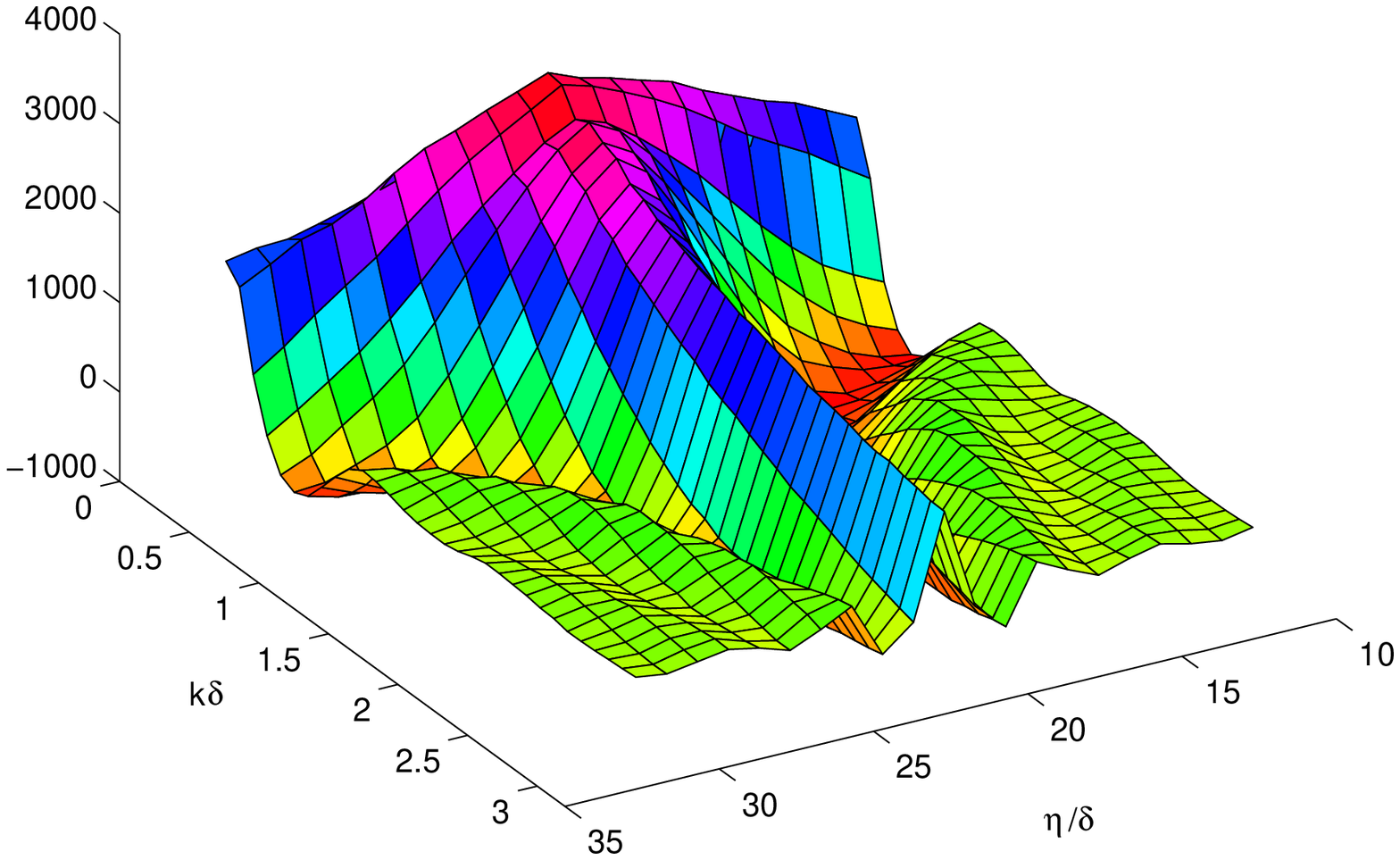,width=7.5cm,angle=0}}
\caption{Two-time correlation function $\TTmom$ for 
$\eta'=22\delta$}
\label{fig:ttcf0i}
\end{figure}

Let us examine how this might work, for superhorizon modes ($k\eta \ll 1$).
Firstly, we divide $\Th_{00}$ and $\Th$ into coherent and incoherent 
parts:
\ben
\Th_{00} = \bar{\Th}_{00} + \De_{00}, \qquad \Th = 3w_s\bar{\Th}_{00} + r,
\een
where $r$ is a random variable, whose fluctuations are limited only 
by the requirement of scale invariance. Thus
\ben
\langle r(\eta_1)r^*(\eta_2)\rangle = \frac{1}{\sqrt{\eta_1\eta_2}} 
R(\eta_1/\eta_2).
\een
It is then not hard to show that energy-momentum conservation 
implies that 
$$
\langle |\De_{00}|^2\rangle \sim \langle |r|^2\rangle.  
$$
Since $\langle |r|^2\rangle < \langle |\Th|^2\rangle$, the extra 
fluctuations in the density induced by the pressure term are 
controlled by the size of the pressure term.  Now, in the 
Minkowski space simulations, we find that $\langle |\Th|^2\rangle 
\sim 0.1 \langle |\Th_{00}|^2\rangle$ for $k\eta \ll 1$, and thus 
we argue that any extra fluctuations in Friedmann models, with a 
coherence time set by the horizon $\eta$, are 
likely to be small.

\section{Conclusions}
We have taken the first steps on the road to calculating the 
matter and radiation power spectra in the cosmic string scenario, 
by constructing a realistic model of the important parts of 
the string energy-momentum tensor which drive the fluid perturbations. 
We have argued that the Minkowski space simulations we used to 
construct the model incorporates the essential features of string 
evolution in an expanding universe, and that corrections due to the 
expansion are small.  This of course should be checked with FRW 
string codes.  Perhaps the most important feature of the string 
network is its coherence time, the time over which the phases of 
the Fourier components of the string energy-momentum tensor are 
correlated.  We find that there is no single 
coherence time for the whole network.  Although the 
fall-off away from the equal-time 
value is modulated by functions whose dominant behaviour 
is $\exp(-(\eta_1-\eta_2)^2/2\eta_c^2)$, and the scale is set by 
$k^{-1}$ for each mode, $\eta_c$ is different for different 
correlation functions.

\acknowledgements{
GRV and MBH are supported by PPARC, by studentship number 
94313367, Advanced
Fellowship number B/93/AF/1642 and grant number GR/K55967. MS is 
supported by the Tomalla Foundation. Partial support is
also obtained from the European Commission under the Human Capital 
and Mobility programme, contract no. CHRX-CT94-0423.}

\vfill
\end{document}